\documentclass[a4paper]{jpconf}
\usepackage{graphicx}
\usepackage{hyperref}
\usepackage{mathtools} 
\usepackage{amsmath}
\usepackage{amssymb}
\usepackage{amsthm}
\usepackage{mathrsfs}
\usepackage{dsfont}
\usepackage{xcolor}

\begin{document}
\title{On the Resilience of Black Hole Evaporation: Gravitational Tunneling through Universal Horizons}

\author{M. Schneider$^{1,2,3}$, F. Del Porro$^{1,2,3}$, M. Herrero-Valea$^4$, S. Liberati$^{1,2,3}$}
\ead{mschneid@sissa.it, fdelporr@sissa.it, mherrero@ifae.es, liberati@sissa.it}

\address{$^1$ Scuola Internazionale Superiore di Studi Avanzati (SISSA),
Via Bonomea 265, 34136 Trieste, Italy}
\address{$^2$ Institute for Fundamental Physics of the Universe (IFPU), Via Beirut 2, 34014 Trieste, Italy}
\address{$^3$ Istituto Nazionale di Fisica Nucleare (INFN) - Sezione di Trieste,
Via Valerio 2, 34127 Trieste, Italy}
\address{$^4$ Institut de Fisica d'Altes Energies (IFAE), The Barcelona Institute of Science and Technology, Campus UAB, 08193 Bellaterra (Barcelona) Spain}

\begin{abstract}
Using a quantum tunneling derivation, we show the resilience of Hawking radiation in Lorentz violating gravity. In particular, we show that the standard derivation of the Hawking effect in relativistic quantum field theory can be extended to Lorentz breaking situations thanks to the presence of universal horizons (causal boundaries for infinite speed signals) inside black hole solutions. Correcting previous studies, we find that such boundaries are characterized by a universal temperature governed by their surface gravity. We also show that within the tunneling framework, given the  pole structure and the tunneling path, only a vacuum state set in the preferred frame provides a consistent picture. Our results strongly suggest that the robustness of black hole thermodynamics is ultimately linked to the consistency of quantum field theories across causal boundaries. 
\end{abstract}

\section{Introduction}
The Hawking effect represents one of the most outstanding discoveries related to black holes~\cite{haw75}, and provides a first glimpse of a deep interconnection between gravity and quantum physics which seems to lead the way into quantum gravity. First introduced by studying the case of the event horizon of a static Schwarzschild black hole, the paradigm of Hawking radiation (or more accurately the Hawking effect) was substantiated \cite{har83} in follow up investigations, and shown to also apply to different sorts of horizons \cite{gib93}, even dynamical ones, cf. \cite{hay94,ash99,ash00,ash03,Barcelo:2010xk,Barcelo:2010pj}. It seems that particle creation near causal boundaries is an inherent feature of quantum field theory in curved space-time. 

However, one might question the dependence of this effect on the properties of the system under consideration, in particular on the presence of Lorentz invariance. Consequences arising from quantum gravity may shake its foundation when obliterating Lorentz symmetry and the merits that come with general relativity. The question that we want to pursue in this article is: ``How resilient is the Hawking effect in absence of Lorentz symmetry?".

In his original article, Hawking pictured the process of particle creation from black holes through a breakup of a virtual pair of particles by the black hole tidal force, leading to a pair production from vacuum where the positive energy particle escapes the gravitational potential, while the negative energy one tunnels through the horizon into the interior. Nonetheless, it was later realized that a complementary picture -- where particles breaks apart close but behind the horizon and the positive energy one tunnels out -- can be advanced. The combinations of these two possible histories led to the so called ``tunneling framework", developed by Parikh and Wilczek \cite{par00}, and also by Padmanabhan and Srinivasan \cite{apd00}. Later on, Massar and Parentani \cite{mas00} applied the Hamilton--Jacobi formalism to the tunneling picture, which opened up the opportunity to study a vast variety of physical systems (see e.g.~\cite{pad02,van08,di09,van11,far14}). This framework also provides a concise description of the Hawking effect regardless of the causal boundary's specifics \cite{gia20}. 

The tunneling picture has turned out to be especially versatile, since it offers a quasi-local framework, even capable of treating mildly dynamical situations. In principle, the idea is to indulgently relate gravitational tunneling through a horizon with tunneling through a potential barrier in quantum mechanics; although in the gravitational case, the field creates the barrier by itself at the horizon \cite{par04}. Although both examples use the WKB method, due care has to be given to both the tunneling path as well as the definition of the observer in the gravitational case.

In the context of general relativity, this framework allows for a universal definition of the Hawking effect which is intimately tied to the internal consistency of quantum field theory across causal boundaries\footnote{By causal boundary we mean here the border of a region of spacetime causally disconnected from the past or the future causal development of its exterior.} \cite{gia20}. For matter characterized by Poincar\'e invariance, i.e. obeying special relativity, the speed of light sets the limit for the propagation of information, and space-time regions can be classified into untrapped (or normal), past or future trapped, and marginally trapped regions by using null-congruences \cite{sen07}. A trapped region is characterized through both null expansions, ingoing and outgoing, being negative. This is tantamount to say that there will be no classical outgoing path available for leaving such a region, and no point of this region can lie in the past of an outside observer. Nevertheless, as shown in e.g. \cite{par00}, paths along complex momenta are still able to cross the horizon in the outgoing direction, leading eventually to a non-zero tunneling rate. 

The notion of a causal barrier is determined by the underlying causal structure, which is dictated by the lightcone and fundamentally by the speed of light, or the utmost speed of information propagation within the system. For general relativity, the Hawking effect is unquestionable, especially within the tunneling picture. Its generalization is straightforward and applies to all types of dynamical horizons \cite{gia20}. The aim of this article is to further extend the application of the tunneling framework to Lorentz violating theories -- in particular to Ho\v{r}ava-Lifshitz gravity \cite{Horava:2009uw} -- which involve dispersion relations with multiple spatial derivatives, allowing for propagation speeds higher than the speed of light, and even infinite speed signals, at high energies \cite{berg12}. Let us note that while departures from Lorentz symmetry are very strongly constrained for mass dimension three, four and five operators extending the standard model of particle physics, very weak constraints are at the moment present for mass dimension six and higher, CPT even, Lorentz breaking operators such as those characterizing Ho\v{r}ava-Lifshitz gravity (see e.g.~\cite{Liberati:2013xla, Addazi:2021xuf}).

Obviously, superluminal signals can cross the standard horizon freely in both directions, as it only obstructs signals that propagate relativistically. It is natural to question then what happens to horizon thermodynamics. To unravel this question, we will have to dig deeper into the structure of Lorentz violating theories, where so called universal horizons, i.e. horizons for infinite speed signals, occur. The existence of the Hawking effect in their presence has been shown in the past literature, cf. \cite{ber13,mi15,he21,fdp} for details. In this article we reinforce this conclusion. In the next section we shall start by reviewing the tunneling framework in relativistic setups, we shall then introduce the so called ``khronometric theory of gravity" i.e.~the so called Einstein-Aether theory supplemented by the condition of having a hypersurface orthogonal aether (which can also be seen as the low-energy limit of Ho\v{r}ava-Lifshitz gravity) in the third section. Then, we derive explicitly the Hawking effect within a generalized tunneling framework for universal horizons, and afterwards we discuss some issues that have been overlooked in the literature so far. Throughout the article, we will work in mostly plus signature.

\section{Gravitational Tunneling: The Relativistic Case}

We consider a spherically symmetric spacetime $(\mathcal{M},g)$ with semi-Riemannian manifold $\mathcal{M}$ and metric $g$ given in the ingoing Eddington-Finkelstein-Bardeen (EFB) coordinate chart via
\begin{equation}\label{bardeen}
g=-e^{2\psi(v,r)}C(r)\mbox{d}v\otimes\mbox{d}v
+2e^{\psi(v,r)} \mbox{d}v\otimes\mbox{d}r
+r^2\mbox{d}\mathbb{S}_2
\end{equation}
with d$\mathbb{S}_2=$d$\vartheta\otimes$d$\vartheta+\sin^2(\vartheta)$d$\varphi\otimes$d$\varphi$ denoting the two-sphere's metric, while $C$ and $\psi$ functions that encode the properties of space-time, and in particular the horizon. Note that these coordinates come in handy to describe black holes and crunching cosmologies, while for white holes and expanding cosmologies one uses instead outgoing EFB coordinates. 

For the simple case of a Schwarzschild black hole of mass $M$, one has $\psi(v,r)\equiv0$ and $C(r)=1-\frac{2M}{r}$. The radius at which $C=0$ marks the position of the horizon \cite{hay09}. For a general analysis, also in dynamical space-times, we refer to \cite{gia20}, while we here focus on Schwarzschild black holes. In this case, \eqref{bardeen} becomes stationary (actually static) and admits a timelike Killing vector field $\chi\propto\partial_v$. For dynamical situations or different symmetries, one needs a generalized concept of the Killing vector \cite{sen15} defining a preferred energy\footnote{In general space-times, one may construct a dual-expansion vector that uses the null-congruences which for spherically symmetric cases reduces to the Kodama vector, and in stationary cases to the Killing vector \cite{sen15}.}.

Gravitational tunneling incorporates the idea that quantum effects across horizons can be described through a complex path crossing the horizon in the geometrically forbidden direction. To formalize this, we employ a WKB approximated massless scalar field $\phi$
\begin{equation}\label{phians}
\phi=\phi_oe^{\frac{i}{\hbar}\sum_p\hbar^p\mathcal{S}_p}\sim\phi_oe^{\frac{i}{\hbar}\mathcal{S}_0}\quad\mbox{for }\hbar\to0,
\end{equation}
where $\mathcal{S}=\sum_p\hbar^p\mathcal{S}_p$ is the scalar field full action, $\mathcal{S}_0$ is the classical action and $\phi_o\in\mathbb{R}=const$ although in principle a mild time-dependence can be allowed as long as the WKB approximation stays valid. Hereinafter, we set $\hbar\equiv 1$. Our scalar field obeys the equation of motion $\Box\phi=0$ with $\Box=g^{ab}\nabla_a\nabla_b$ being the d'Alembert operator. Applying this differential operator to \eqref{phians} yields an equation for the classical action $\mathcal{S}_0$ -- also called principal function -- which is solved by the ansatz $\mathcal{S}_0=\int\mbox{d}x^ak_a$ with $k_a=\partial_a\mathcal{S}_0$ being the four-momentum. This object describes a point particle action corresponding to our picture of a particle crossing the horizon\footnote{This monochromatic wave approximation only holds close to the horizon as a consequence of the stationary phase approximation. If we instead wanted to trace its propagation path, we would need to integrate over the frequency.}. For the case of a future outer horizon\footnote{For the full classification of dynamical and static horizons, cf. \cite{hay94,ash99}.}, i.e.~a black hole like horizon, we then find the principal function $\mathcal{S}_0$ for a space-time with metric \eqref{bardeen} to be
\begin{equation}\label{action}
\mathcal{S}_0=-\int \Omega \mbox{d}v +\int\mbox{d}rk_r(r)+W(\sphericalangle),
\end{equation}
where $W(\sphericalangle)$ is the angular contribution to the action and, for the $v$-component, we used the Killing energy $\Omega:=-\mathcal{L}_\chi\mathcal{S}_0=-\chi^a\partial_a\mathcal{S}_0=-e^{-\psi(v,r)}\partial_v\mathcal{S}_0$ associated to the Killing vector field $\chi_a$. From now on, we restrict the analysis to s-waves, such that the angular contribution $W(\sphericalangle)\equiv0$. Using \eqref{action}, we find the Hamilton--Jacobi equation for $k_r$ and $\Omega$ to be solved by 
\begin{equation}\label{radialimpuls}
k_r(r)=\frac{2\Omega}{C(r)},
\end{equation}
which develops a pole in the d$r$-integral at the position of the horizon where $C(r)=0$. This pole will turn out to be essential for the tunneling picture as we show below. 

Recollecting the quantum mechanical tunneling through a barrier, the tunneling rate $\Gamma$ is derived by a comparison between the incident and the transmitted wave function
\begin{equation}\label{tunneling_rate}
\Gamma=\frac{\|\phi_{\rm trans}\|^2}{\|\phi_{\rm inc}\|^2}\propto e^{-\frac{2}{\hbar}{\rm Im}(\mathcal{S}_0)}
\end{equation}
where $\|\phi\|$ denotes the $L^2$-norm. For gravitational tunneling, the rate is defined equally, such that the above formula holds as well. This formula unveils that the tunneling is connected to an eventual imaginary part of $\mathcal{S}_0$, which in turn can be linked to the entropy. According to \eqref{action}, the tunneling occurs along $k_r(r)$. This is the direction of horizon crossing provided we face a spherically symmetric setup, as we do.
\begin{figure}
\includegraphics[width=13pc]{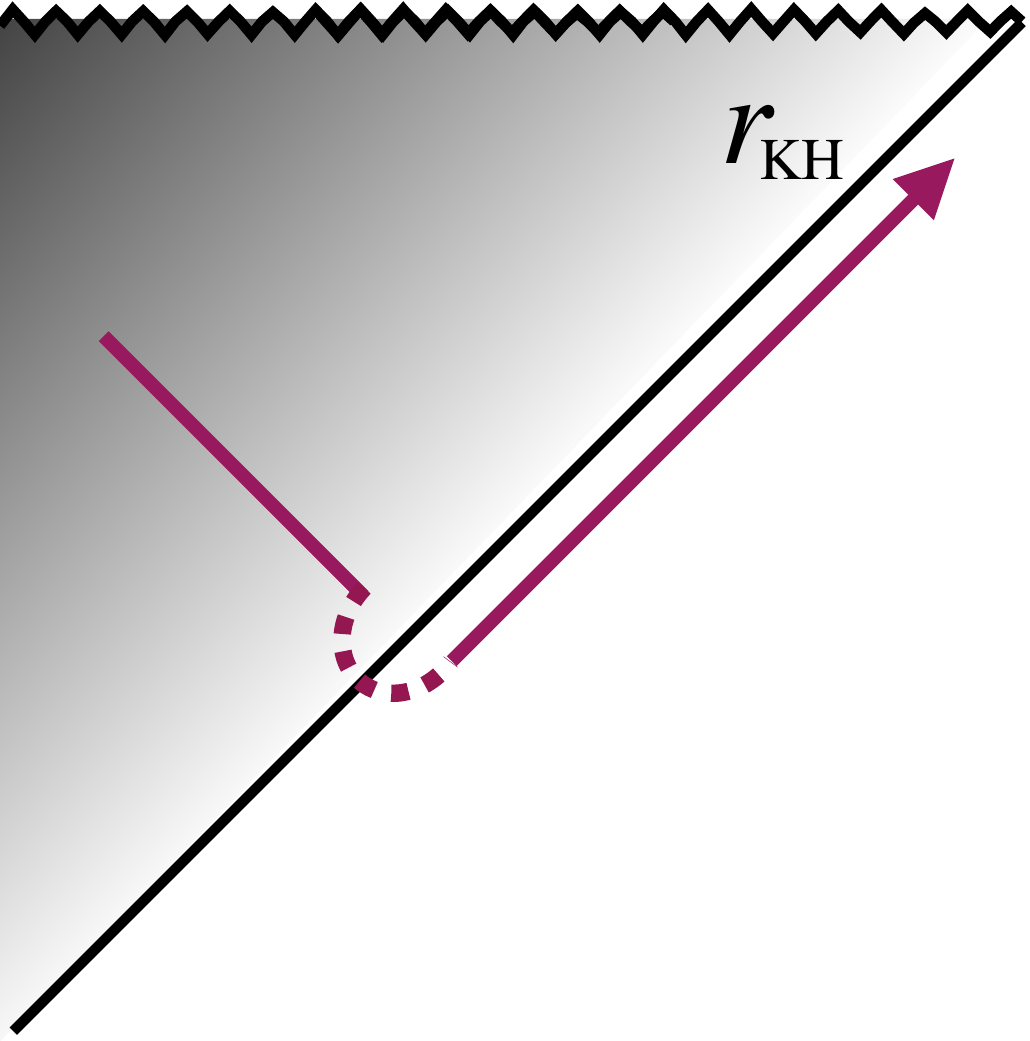}\hspace{7pc}%
\begin{minipage}[b]{14pc}\caption{\label{grtunath}Tunneling path for a relativistic black hole at the radius $r_{\rm KH}$. The horizon for static black holes is a null surface on which the Killing vector vanishes, i.e. a Killing horizon. The shaded area depicts the trapped region and the zig-zag line the singularity at $r=0$. Straight lines represent classical paths while the dashed line describes the complex path on which the tunneling occurs.}
\end{minipage}
\end{figure}

In fact, the pole in \eqref{radialimpuls} introduces a non-trivial imaginary part. By expanding around the horizon $C(r)\simeq \kappa_H(r-r_H)+\mathcal{O}((r-r_h)^2)$ where we assumed $C(r)\in\mathcal{C}^2(\mathbb{R})$ at least and defined $\kappa_H:=\frac{\partial C}{\partial r}\vert_{r_H}$ to be the horizon's surface gravity. To see the influence of the pole explicitly, we introduce an infinitesimal complexification through an $i\varepsilon$-prescription, such that $C(r)\to C(r+i\varepsilon)$ and find the imaginary part to be independent of $\varepsilon$ (cf. \cite{gia20} for details)
\begin{equation}
{\rm Im}(\mathcal{S}_0)={\rm Im}\left(\int_{r_1}^{r_2}\frac{\Omega{\rm d}r}{\kappa_H(r-r_H-i\varepsilon)}\right)=\frac{\pi\Omega}{\kappa_H},
\end{equation}
where the tunneling path is chosen to be radially outgoing from $r_1$ to $r_2$ while $r_1<r_H<r_2$, and null at a fixed value in the $v$-direction. For a Schwarzschild black hole, the surface gravity $\kappa_H=1/4M$ ergo Im$(\mathcal{S}_0)=4\pi M\Omega$, which is the well-known result from performing an analytic continuation across the horizon \cite{Jacobson:2003vx}. 

The connection of the above result with thermodynamics comes by introducing a set of observers with respect to which one can define a concept of vacuum and measure the particle content in a given state. In particular, assuming a set of coordinates and the existence of a vacuum state both well behaving across the horizon, one can compare the tunneling rate \eqref{tunneling_rate} with a Boltzmann distribution \cite{har83}. Therefore, whenever Im$(\mathcal{S}_0)\propto\Omega$, the observer can read off a horizon temperature
\begin{equation}\label{boltz}
\Gamma=e^{-2{\rm Im}(\mathcal{S}_0)}\triangleq e^{-\frac{\Omega}{T_H}},\quad\mbox{whence}\quad T_H=\frac{\kappa_H}{2\pi}.
\end{equation}

Notice that the definition of the observer is critical to extract such a temperature. In particular, the vacuum state is constructed in a local frame which is determined by the vector field defining the notion of energy. Note also that the above reasoning readily applies also to dynamical horizons as long as the WKB approximation, $\kappa_H\ll \Omega$, holds.

Let us comment on \eqref{boltz} a bit more: the comparison with the Boltzmann factor ultimately relates thermodynamics with a consistent quantum field theory across the horizon. In particular, the tunneling picture provides a definition of the {Hawking effect} \cite{gia20} that is given by the very handy expression
\begin{equation}\label{ims0}
\mbox{Im}(\mathcal{S}_0)>0
\end{equation}
which guarantees a positive definite temperature. 

The same property provides an implicit argument in support of the conclusion that the quantum field theory is consistent because it keeps the probabilistic interpretation intact. To see this, we consider the origins of the tunneling formalism within the lore of quantum field theory (QFT). Despite its simplicity, the tunneling formalism's principles are rooted within the Hadamard bi-distribution. Let us only give a brief argument while a more sophisticated reasoning can be found in \cite{mor12,kur21}. In these references, it is shown that within the quasilocal approximation, the Hadamard two-point bi-distribution across the horizon behaves as $\langle\phi(r_2)\phi(r_1)\rangle\sim\Gamma\sim e^{-\Omega/T}$ which immediately implies finiteness and a sensible probabilistic interpretation as long as $T>0$. In the opposite case, the probabilistic interpretation of the theory would be violated fundamentally. Due to this, horizon thermodynamics serves as a beacon for the consistency of quantum field theory across causal boundaries.

Having reviewed the tunneling derivation of the Hawking effect, let us now move to describe the gravitational setting that we shall adopt in our extension of the tunneling derivation.


\section{Lorentz-violating Gravity} 

A fully consistent framework for gravity beyond local Lorentz invariance is given by Ho\v{r}ava--Lifshitz gravity~\cite{Horava:2009uw}. This is an extension of General Relativity characterized by the presence of a preferred foliation orthogonal to a timelike and hypersurface orthogonal (irrotational) vector field -- the aether. This allows for the construction of a renormalizable theory of gravity by the addition of spatial higher derivatives to the action (see~\cite{Anselmi:2007ri,Barvinsky:2015kil,Barvinsky:2017kob,Barvinsky:2019rwn,Bellorin:2022qeu} for promising investigations concerning perturbative renormalizability and \cite{Barvinsky:2023mrv} for a review).

In four space-time dimensions, Ho\v{r}ava--Lifshitz gravity is characterized by CPT even, mass dimension four, six and eight operators, however low energy classical black hole solutions relevant for this study are assumed to be determined only via the low energy (mass dimension four operators) part of the action. This was found to coincide with the so called Einstein--Aether gravity~\cite{Jacobson:2000xp} once hypersurface orthogonality is imposed on the aether field. In this case, the aether field contains only a single scalar degree of freedom, which takes the name of khronon, and the theory is referred to as ``khronometric gravity". Surprisingly, staticity and spherical symmetry automatically impose this condition, which must be added \emph{a posteriori} in more general settings.

As a consequence of the introduction of the aether, the doublet $(\mathcal{M},g)$ becomes a triplet $(\mathcal{M},g,U)$ where $U$ is the \emph{aether} one-form\footnote{In fact, even in relativistic systems, the space-time is a triplet $(\mathcal{M},g,\mathfrak{o})$. Provided that the manifold is globally time-orientable, we have an orientation $\mathfrak{o}$ which in principal can coincide with $U$. Nevertheless, the vector pointing along the orientation in general relativity is not dynamical, whilst the aether is so \cite{Jacobson:2000xp}.}. Once the aether is fixed, and since it defines a preferred foliation, the gravitational symmetry group of diffeomorphisms $\mathfrak{Diff}(\mathcal{M})$ reduces to the group of foliation preserving diffeomorphisms $\mathfrak{FDiff}(\mathcal{M})$
\begin{equation}
t\to t'(t)\quad\mbox{(time reparametrization)}\quad\mbox{and}\quad x\to x'(t,x)\quad\mbox{(spatial diffeomorphism)}.
\end{equation}

The action of Einstein--Aether theory consists of the Einstein-Hilbert action combined with a normalized timelike aether such that
\begin{equation}
S_{\rm EAG}=\int_\mathcal{M}\mbox{d}^4x\sqrt{-g}\left(\mathcal{R}+{K_{ab}}^{cd}\nabla_cU^a\nabla_dU^b-\lambda( U^a U_a+1)\right)
\end{equation}
where $\mathcal{R}$ is the Ricci scalar and $\lambda$ is a Lagrange parameter implementing the unit norm condition for $U$. The tensor $K$
\begin{equation}
{K_{ab}}^{cd}=c_1g_{ab}g^{cd}+c_2{\delta_a}^c{\delta_b}^d+c_3{\delta_a}^d{\delta_b}^c+c_4U^cU^dg_{ab}
\end{equation}
contains four coupling constants $c_i\in\mathbb{R}$ which determine the specifics of the aether \cite{Jacobson:2000xp}. Their parameter space is tightly constrained mostly by strong gravity observations, but a sufficient large island compatible with observations remains allowed (see e.g.~\cite{Mattingly:2005re, yagi14,Gupta:2021vdj}). 

For some unequivocal combinations of the coefficients, black hole like solutions are known analytically, cf. \cite{eli06, bar11} for details. In our analysis, we consider a Schwarzschild-like metric of the form
\begin{equation}
g=-C(r)\mbox{d}v\otimes\mbox{d}v+2\mbox{d}v\otimes\mbox{d}r+r^2\mbox{d}\mathbb{S}_2.
\label{staticMet}
\end{equation}

Concerning the aether, staticity and spherical symmetry enforce hypersuface orthogonality automatically. In such case, $U$ is fully determined by a scalar function $\varsigma(x)$, aptly dubbed \emph{khronon}
\begin{equation}
U_a=\frac{\partial_a\varsigma(x)}{\|\partial \varsigma\|_2}.
\end{equation}
Hence, the aether defines a preferred time-direction on the full space-time $(\mathcal{M},g,U)$ and has to obey the same isometries of space-time. This implies the form 
\begin{equation}
U=-\frac{1+C(r)A^2(r)}{2A(r)}\mbox{d}v+A(r)\mbox{d}r,
\end{equation}
where $A(r)$ is an aether exclusive function.

Given $U$, it is possible to introduce its normal spatial $S$, that is, the vector always tangent to the foliation leafs.
This can be readily deduced as 
\begin{equation}\label{es}
S=\frac{1-C(r)A^2(r)}{2A(r)}\mbox{d}v+A(r)\mbox{d}r
\end{equation}
which is normalized $\|S\|=1$ and chosen to be outward pointing. 
As can be checked easily, the form in \eqref{es} ensures orthogonality with the aether $U$. 

Provided these two vectors we can identify a corresponding preferred frame $(\tau,\rho,\vartheta,\varphi)$ spanned by $U$, $S$, as well as the angular part, such that the time $\tau$ aligns with the aether: $U_a$d$x^a=U_\tau$d$\tau$ while the spatial coordinate $\rho$ is defined through a similar equation involving $S$. 
We can relate the $(v,r)$ system of coordinates to the $(\tau,\rho)$ one by simple relations
\begin{equation}\label{tau}
\tau=v+\int\frac{U_r}{U_v}\mbox{d}r\quad\mbox{and}\quad\rho=v+\int\frac{S_r}{S_v}\mbox{d}r.
\end{equation}

The preferred coordinate system $(\tau,\rho,\vartheta,\varphi)$ can be used to recast the metric \eqref{bardeen} into the ADM or $(1+3)$-form by introducing the lapse function $N$, the shift vector $N^a$ and the induced metric $\gamma=g-U\otimes U$ on the spatial submanifolds. The metric reads
\begin{equation}\label{rho}
g=-(N^2-N^aN_a)\mbox{d}\tau\otimes\mbox{d}\tau+N_a\mbox{d}x^a\otimes\mbox{d}\tau+\gamma_{ab}\mbox{d}x^a\otimes\mbox{d}x^b\, .
\end{equation}
The choice of $\rho$ above, which is not unique and can be changed by a spatial diffeomorphism, corresponds to the gauge choice $N^a=0$, which is always possible. Note, although the angular part d$\mathbb{S}_2$ remains unchanged, its prefactor $r^2=r(\tau,\rho)^2$ can now be time-dependent. 

Having covered the gravity part, let us now briefly review the matter sector and introduce the Lorentz breaking action of a massless and real  Lifshitz scalar field 
\begin{equation}\label{action_lifshitz}
S_{\rm LSF}=-\frac12\int{\rm d}^4x\sqrt{-g}\left[\partial_a\phi\partial^a\phi+\sum_{j=2}^n\frac{\alpha_{2j}}{\Lambda^{2j-2}}\phi(-\Delta)^j\phi\right]
\end{equation}
where $\Lambda$ is the energy scale associated to Lorentz breaking, $\alpha_{2j}\in\mathbb{R}$ with $\alpha_{2n}\equiv1$, and $\Delta=\gamma^{ab}\nabla_a\nabla_b$ if the Laplace operator restricted to the foliation leafs \cite{Pospelov:2010mp}. In the preferred frame $\Delta$ is a purely spatial operator and hence the action remains explicitly second order in time derivatives and thus free of Orstrogradsky instabilities. This conclusion must be the same in any other frame of coordinates, albeit not apparent~\cite{Blas:2010hb}. In terms of model building, one can think of this action as either a toy model for Lorentz violating perturbation on top of this geometry -- for instance, pure gravitational perturbations, whose direct study remains an open challenge -- or, alternatively, as the action of an external scalar field coupled to gravity. In the latter case, the scale $\Lambda$ has to be sufficiently high to fulfill observational and experimental constraints, or a mechanism to suppress these operators must be provided \cite{Pospelov:2010mp}. The maximum number of spatial derivatives $2n$ is usually chosen to achieve power-counting renormalizability. In the case of a gravitational action, this leads to $n=3$ in four space-time dimensions. Here however, we leave $n$ arbitrary for the sake of generality.

Additionally, it can be shown that \eqref{action_lifshitz} allows for superluminal propagation and as such special attention must be paid to identifying the causal boundaries of our space-time. Obviously, this property changes the causal structure of these theories fundamentally \cite{Bhattacharyya:2015gwa}. Provided one allows infinite speed signals, timelike connected points are all points on future leafs while the notion of spacelike separation is solely related to points on the same hypersurface; the notion of lightlike is hence obsolete. 

Nevertheless, even within this framework, trapped regions can be rigorously defined~\cite{Carballo-Rubio:2020ttr,Carballo-Rubio:2021wjq}. Similarly, the notion of stationary black hole can be extended~\cite{eli06,bar11}: for $(\mathcal{M},g,U)$ being a Schwarzschild-like space-time, the asymptotic boundary at spatial infinity $\mathfrak{i}^0$ has to be flat, i.e. $C(r)\to1$ and $A(r)\to1$ for $r\to\infty$, while a singularity is lurking at the origin. To ensure smooth asymptotics, the product $U\cdot\chi\to -1$ in this limit, thus infinite speed signals reach $\mathfrak{i}^0$. 

Note that we have defined the contraction between one-form and vector as usual, and that $U\cdot\chi=U^a\chi_a=-N$. Having a trapped region for those signals implies the existence of leafs that are unable to fulfill the above given asymptotic boundary conditions, or in other words, they must be compact spatial submanifolds, and as such not reaching $\mathfrak{i}^0$. With this knowledge, a \emph{universal horizon} is defined as the constant radius surface (to maintain spherical symmetry) at which \cite{Bhattacharyya:2015gwa}
\begin{equation}
U\cdot\chi=0\quad\mbox{and}\quad A\cdot\chi\neq0,
\end{equation}
where $A^a=\mathcal{L}_UU^a=U^b\nabla_bU^a$ is the aether acceleration. 

Let us emphasize, that if and only if for a submanifold (at radius $r_{\rm UH}$ in spherically symmetric setups) both of the above conditions are satisfied, then this particular leaf is the outermost compact submanifold, thus being a horizon. Moreover the surface of radius $r_{\rm UH}$ bears the property of being a surface of simultaneity, as well as a sphere of symmetry. Since this leaf can never reach spatial infinity, no signal with infinite speed - straddling the leaf - can escape and is hence trapped. Although the Killing horizon does not provide an ultimate separation principle for Lifshitz fields, the universal horizon fills in and becomes an ideal candidate to possess thermodynamical properties. 

\section{Gravitational Tunneling: The Non-Relativistic Case}

Since the universal horizon provides the space-time with a causal boundary, we can perform the tunneling analysis in its neighborhood, taking care of various subtleties that arise within these theories. Near the universal horizon, the WKB approximation holds and allows to write the principal function \eqref{action} in the preferred frame as (again for an s-wave) \cite{ber13,fdp}
\begin{equation}
\mathcal{S}_0=-\int\omega U_a\mbox{d}x^a+\int k_\rho S_a\mbox{d}x^a.
\end{equation}
Here we have defined $\omega=-U\cdot k$ and $k_\rho=S\cdot k$ using the covariant momentum $k=\omega U+k_\rho S$ in the aether frame. Plugging this into the WKB approximation for $\phi$, we derive the Hamilton--Jacobi equation through application of $\Box-\sum_{j=2}^n\alpha_{2j}(-\Delta)^j/\Lambda^{2j-2}$ onto the field. Hence, we find\footnote{The Hamilton-Jacobi equation seems different with respect to the one in \cite{fdp}. This results from taking a different ansatz for $\mathcal{S}_0$ involving the lapse in the definition of $k_a$, cf. \eqref{k0}.}
\begin{equation}\label{hjeq}
-\omega^2+\gamma^{\rho\rho}k_\rho^2+\frac{(\gamma^{\rho\rho})^n}{\Lambda^{2n-2}}\left(k_\rho^{2n}+G(k_\rho,\nabla k_\rho;\Lambda)\right)=0
\end{equation}
where $G(k_\rho,\nabla k_\rho;\Lambda)$ contains polynomial terms of order lower than $2n$, as well as derivatives of $k_\rho$, which are suppressed in the limit of high momenta $k_\rho\gtrsim\Lambda$ within the zeroth order eikonal approximation\cite{bhatt}. 

With respect to the Lorentz-breaking scale, the Hamilton--Jacobi equation develops two distinct regimes, that we can conveniently call \emph{soft regime} ($k_\rho\ll\Lambda$) and \emph{hard regime} ($k_\rho\gg\Lambda$)
\begin{eqnarray}
-\omega^2+\gamma^{\rho\rho}k_\rho^2+\mathcal{O}\left(\frac{k_\rho}{\Lambda}\right)&=&0,\quad(\mbox{soft regime: } k_\rho\ll\Lambda)\label{soft}\\
-\Lambda^{2n-2}\omega^2+\left(\gamma^{\rho\rho}\right)^nk_\rho^{2n}+\mathcal{O}\left(k_\rho^{2n-2}\right)&=&0,\quad(\mbox{hard regime: } k_\rho\gg\Lambda)\label{hard}
\end{eqnarray}
Let us stress, that in our previous analysis~\cite{fdp}, the above mentioned limits \eqref{soft} and \eqref{hard} were performed solely with respect to $\Lambda$ (the soft regime corresponding to $\Lambda\to 0$, and the hard regime to $\Lambda\to \infty$). While this might appear equivalent to the limiting procedure adopted above, we shall see that the present choice is more accurate as does not neglect relevant terms in the near horizon limit which will slightly change the results of this paper with respect of those reported in~\cite{fdp}.

It is also important to keep in mind that in the above expressions, the frequency $\omega$ is not a constant of motion, differently from the Killing frequency $\Omega$. What we know, nevertheless, is that for an asymptotic observer these two frequencies must coincide, i.e.~$\Omega=\omega$ at $\mathfrak{i}^0$. 
Using the definition of the Killing energy, which is still a constant of motion due to staticity of the solution, and its decomposition within the preferred frame, we find
\begin{equation}\label{ee}
k\cdot\chi=\Omega=\omega (U\cdot\chi)-k_\rho (S\cdot\chi).
\end{equation} 

\begin{figure}
\includegraphics[width=20pc]{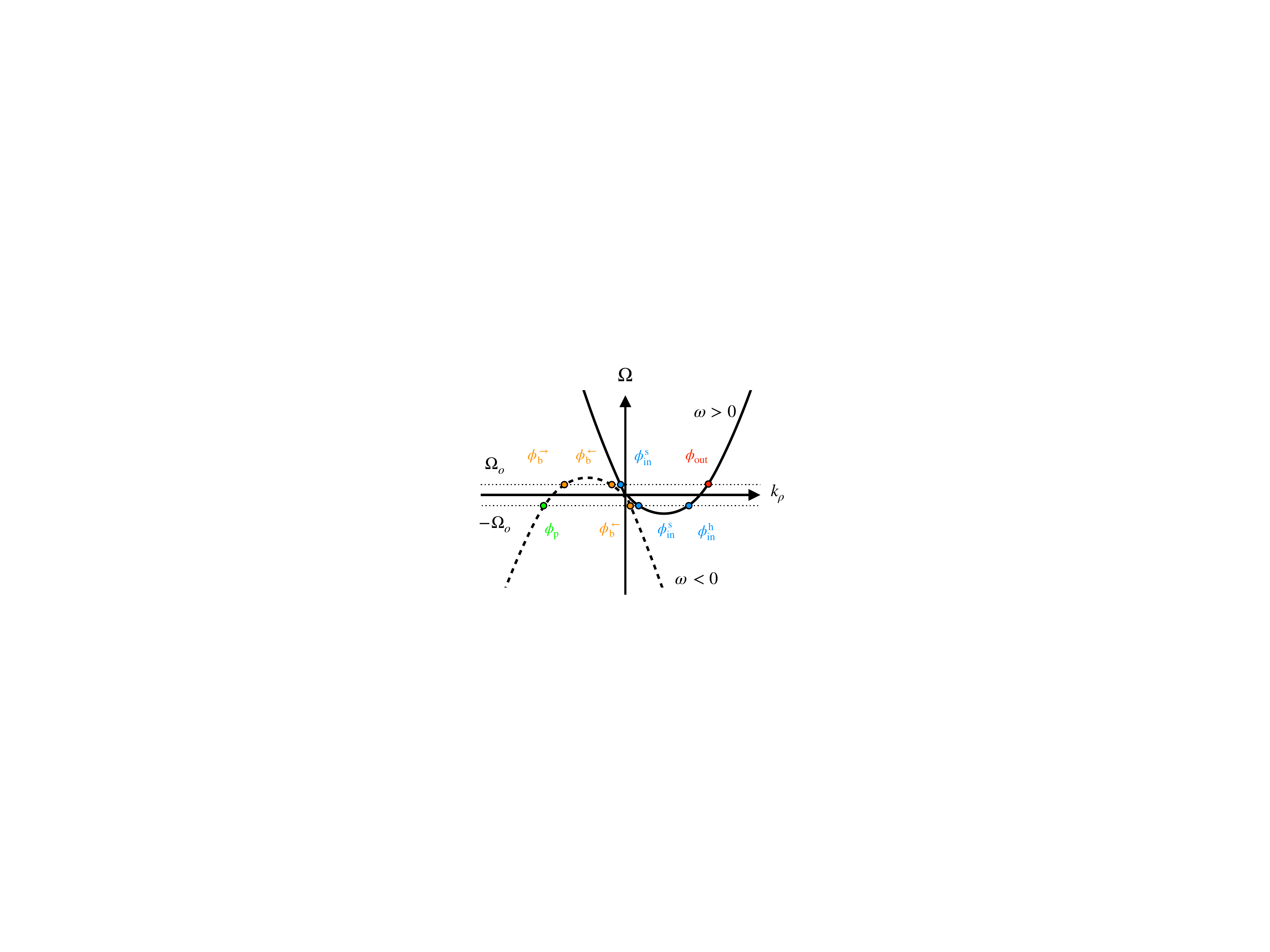}\hspace{1pc}
\begin{minipage}[b]{17pc}\caption{\label{dr} Illustration of the dispersion relation. The two {\em parabolae} of the dispersion relation are shifted such that for a fixed $\Omega_o$ (dotted line) it is possible to see two modes with positive preferred energy $\omega$ (straight line) and two with negative $\omega$ (dashed line). These are the soft modes (dots close to $k_\rho=0$) $\phi_{\rm in}^{\rm s}$ and $\phi_{\rm b}^\leftarrow$, and the hard modes $\phi_{\rm b}^\rightarrow$ and $\phi_{\rm out}$. If we peek beyond the universal horizon, we need to consider a negative $\Omega_o$, and we find again two soft modes $\phi_{\rm in}^{\rm s}$ and $\phi_{\rm b}^\rightarrow$ as well as two hard ones $\phi_{\rm in}^{\rm h}$ and $\phi_{p}$. We refer to figure \ref{modes} for a graphical representation with matching color scheme.}
\end{minipage}
\end{figure}

\begin{figure}
\includegraphics[width=18pc]{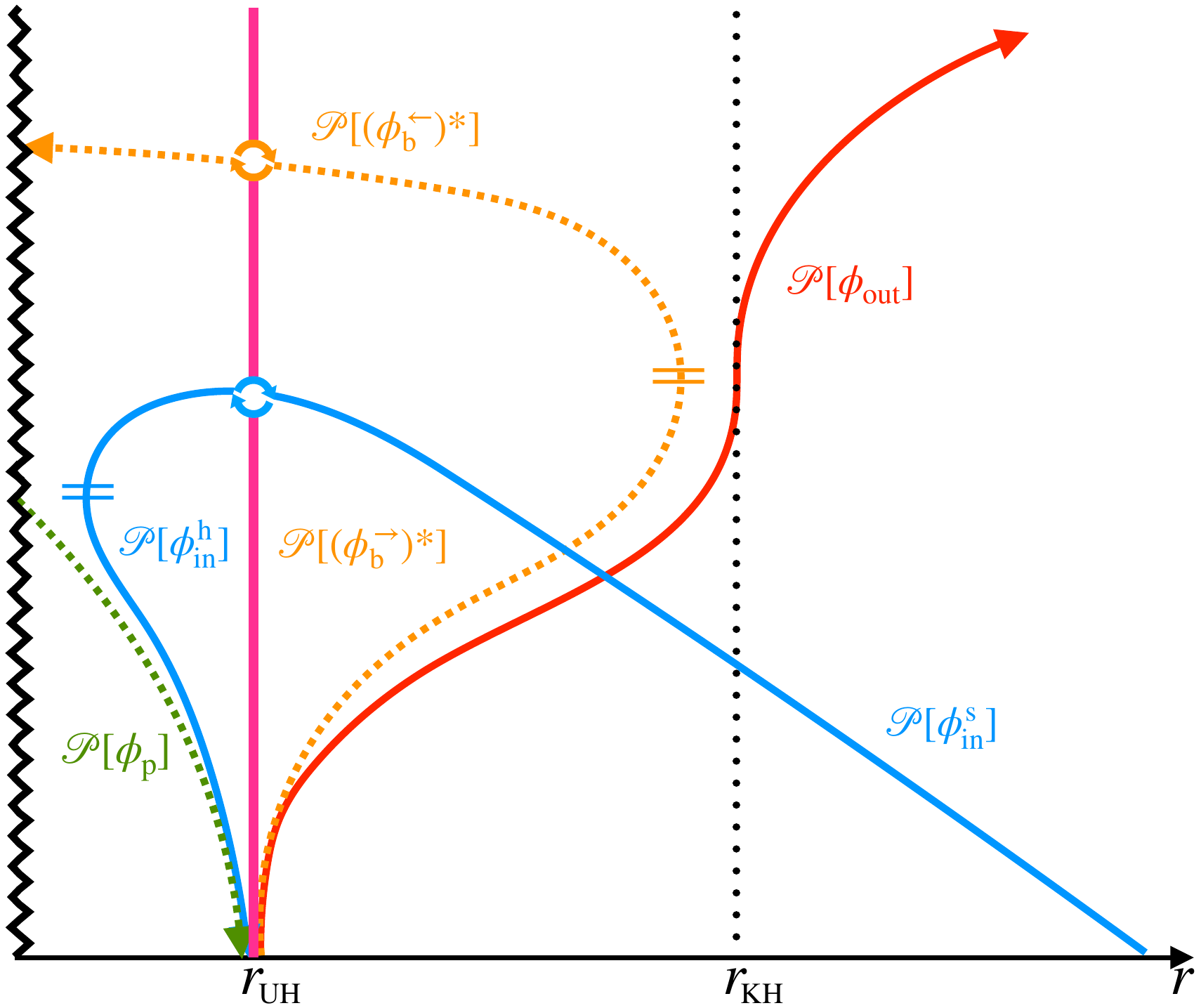}\hspace{2pc}
\begin{minipage}[b]{18pc}\caption{\label{modes}Representation of the wave packets $\mathscr{P}$ created from the real modes: the ingoing (blue) $\phi_{\rm in}$ and the outgoing (red) $\phi_{\rm out}$ positive energy modes, its Hawking partner (green) $\phi_{\rm p}$ as well as the bounded (orange) mode $(\phi_{\rm b})^*$. The latter is conjugated in order to have positive preferred energy oputside the universal horizon. Note, $\phi_{\rm out}$ and $\phi_{\rm p}$ are always hard modes, while $\phi_{\rm in}$ is soft outside (i.e. $\phi_{\rm in}^{\rm s}$) and develops a hard branch $\phi_{\rm in}^{\rm h}$ inside the horizon. Similar for the bound mode, which has hard outbound $(\phi_{\rm b}^\rightarrow)^*$ and soft inbound branches $(\phi_{\rm b}^\leftarrow)^*$. Here, the universal horizon is denoted by the purple line, the Killing horizon as a dotted line and the zigzag line depicts the singularity.
}
\end{minipage}
\end{figure}

In order to better illustrate the underlying dynamics of the system, let us consider for a moment the case $n=2$, for which we can solve the resulting dispersion relation $\omega^2=N^2(k_\rho^2+\alpha_4k_\rho^4/\Lambda^2)$ exactly. After using \eqref{ee}, we can plot this modified dispersion relation close to the universal horizon, as shown in figure \ref{dr}. For a given energy $\Omega_o$, we can find up to four intersection points which represent different propagating modes. Soft modes can be found close to the origin, while hard ones are at larger values of $k_\rho$. We schematically describe the real modes' wave packets as $\mathscr{P}[\phi]=\int$d$\Omega \phi(\Omega)\Upsilon(\Omega)$ where $\Upsilon(\Omega)$ is the energy profile of the corresponding packet in figure \ref{modes}. 
 
The soft modes consist of modes that cross the universal horizon smoothly, i.e. the principal function develops no pole there. We shall label them $\phi_{\rm in}$ and $\phi_{\rm b}$, being respectively the infalling positive energy mode (blue) and the bounded negative energy mode (orange). 

The hard modes instead peel-off the universal horizon, describing the outgoing positive energy mode $\phi_{\rm out}$ (red) and its Hawking partner $\phi_{\rm p}$ (green) that hits the singularity. Note that the bounded mode has two branches, one hard outbound $\phi_{\rm b}^\rightarrow$ mimicking the outgoing mode and a soft inbound $\phi_{\rm b}^\leftarrow$ crossing the universal horizon. The two can be joined into a single trajectory leaving the universal horizon, approaching the interior of the Killing horizon, and then bending inwards to plunge into the universal horizon and the singularity. We will come back to the significance of this set of modes in a later part of the article. Note as well that the same behavior can be observed for $\phi_{\rm in}$ in the interior region.

Now it is time to discuss the tunneling path explicitly. Lorentz breaking theories develop quantum instabilities unless we operate within the preferred frame \cite{Anselmi:2007ri}. As a consequence, the vacuum as well as the choice of the observer are necessarily given by the aether. Similarly, we cannot \emph{a priori} align our spacial momentum with the tunneling path. It should be mentioned that one could perform a rotation into the EFB frame. However, by doing this we only find concordance for the soft modes, where the effect of the Lifshitz term in $S_{\rm LSF}$ is suppressed. For the hard modes instead, pathologies arise due to an essential singularity in $\mathcal{S}_0$, unless the analysis is restricted to the zero mode. 

\begin{figure}
\includegraphics[width=13pc]{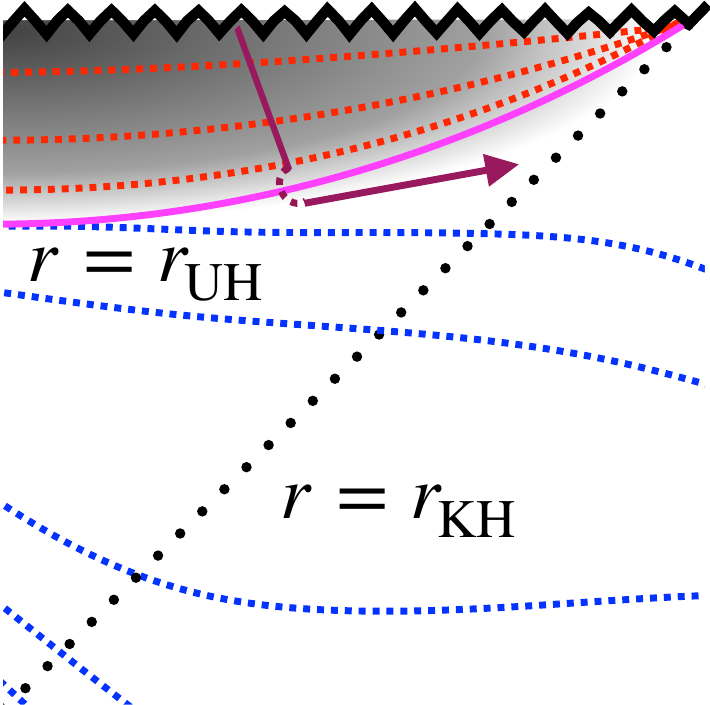}\hspace{7pc}%
\begin{minipage}[b]{14pc}\caption{\label{eatunath}Illustration of the tunneling path for a universal horizon at radius $r_{\rm UH}$ (pink straight line) and Killing horizon at radius $r_{\rm KH}$ (dotted line). The blue dashed lines correspond to the non-compact outer foliation leafs, and the red dashed lines to the compact inner foliation leafs. The magenta line describes the tunneling path, with a complex dashed section.}
\end{minipage}
\end{figure}

The tunneling path for the s-wave will cross the universal horizon perpendicularly, i.e.~in the $r$-direction of the EFB frame. To get the trajectory, we need to introduce an evolution parameter to formalize the tunneling path. A natural and physical choice consists in choosing the preferred time. If the path respects spherical symmetry (as shown in Figure \ref{eatunath}), this boils down to find the relation $\rho=\rho(\tau)$.

To construct the path for the field across the universal horizon explicitly, we use a geometrical optics approximation on top of the WKB description. We thus require ${\rm d}\phi/{\rm d}\tau=0$ -- meaning that the phase of a ray remains constant \cite{cro14}. Explicitly
\begin{align}
    \frac{{\rm d}\phi}{{\rm d}\tau}=\frac{\partial\phi}{\partial \tau}+\frac{\partial\phi}{\partial \rho}\frac{{\rm d} \rho}{{\rm d} \tau},
\end{align}
from which ${\rm d} \rho/{\rm d}\tau$ can be obtained and used to evaluate the principal function on the trajectory along $\tau$. We can start by noticing that in the preferred frame $U_\tau=N$ and $S_\rho=V=(S\cdot \chi)$. Then, using the following relations
\begin{equation}
\partial_\tau\phi=i\omega U_\tau\phi, \quad \partial_\rho\phi=ik_\rho S_\rho \phi, \label{k0}
\end{equation}
we obtain
\begin{equation}
\frac{\mbox{d}\rho}{\mbox{d}\tau}=\frac{N\omega}{Vk_\rho}\label{traj}
\end{equation}
which has to be fed with the specifics of the trajectory, i.e. the behavior of $\omega$ and $k_\rho$. 

Now, for defining the field to be in a vacuum state across the horizon, we need to extend the preferred frame beyond it. Adopting our pathway from the relativistic black hole, we first acknowledge that at $r_{\rm UH}$, the preferred time $\tau\to\infty$ while the lapse $N\to0$. 
However, there is a catch when using $\tau$. Albeit the preferred time's range $\tau\in\mathbb{R}$, it comprises only the exterior region, that is $r\in(r_{\rm UH},\infty)$ while the interior part $r\in(0,r_{\rm UH})$ possesses its own preferred time $\tau'\in\mathbb{R}$. 

Nevertheless, a remedy is presented by promoting the lapse function to a suitable temporal variable. For this, $N(\tau)\in\mathcal{C}^2(\mathbb{R})$ such that $N$ must flip sign in the interior \cite{del22}. Therefore, $N<0$ as perceived by the outside observer. As a consequence, the interior's time decreases towards the singularity and the map $\tau=\tau(N)$ is multi-evaluated. This can be seen clearly by noting that close to the universal horizon $\tau \sim \ln|N|$ (cf.~\cite{del22} for details), which indeed maps two different foliations, for $N>0$ -- corresponding to the exterior time $\tau$ -- and $N<0$ -- the interior time $\tau'$. Using $N$ -- which is tantamount to using the integral lines of $U$ as time coordinate -- we can connect both regions smoothly through the horizon. As a result, close to the universal horizon we have
\begin{equation}\label{taulapse}
\mbox{d}\tau=\frac{{\rm d}r}{N}=\frac{1}{2\kappa_{\rm UH}}\mbox{d}\ln|r-r_{\rm UH}|\simeq\frac{1}{2\kappa_{\rm UH}}\mbox{d}\ln|N|=\frac{1}{2\kappa_{\rm UH}}\frac{\mbox{d}N}{N},
\end{equation}
where we have defined the surface gravity $2\kappa_{\rm UH}=\partial_r N(r_{\rm UH})$ according to the definition of ``peeling surface gravity" given in \cite{cro13}. Hence, close to the universal horizon, the lapse function behaves as $N\propto e^{\kappa_{\rm UH}\tau}$ outside and $N\propto-e^{-\kappa_{\rm UH}\tau'}$ inside the universal horizon respectively, with the different sign due to the lapse's sign flipping \cite{par00,del22}.
The above relation can also be understood as analytically continuing the lapse into the trapped region, similar as in the relativistic case when going from the Boulware to the Unruh state. We are now in the position to analyze the solution space of the Hamilton--Jacobi equation in the two regimes.

\subsection{Soft Regime}
The soft regime is characterized by small momenta compared to the Lorentz-breaking scale, $k_\rho\ll\Lambda$. Higher order terms in the dispersion relation are suppressed by the Lorentz-violating scale and we obtain a simple Hamilton--Jacobi equation that is solved close to the universal horizon by
\begin{equation}\label{kweich}
k_\rho=-\frac{\Omega}{(S\cdot\chi)}
\end{equation}
which shows no pole structure whatsoever. Let us evaluate the principal function's imaginary part Im$(\mathcal{S}_0)$ for the solution \eqref{kweich}, then we find 
\begin{equation}
{\rm Im}\left(\mathcal{S}_0\right)={\rm Im}\left(\int N\omega{\rm d}\tau+\int V k_\rho{\rm d}\rho\right)={\rm Im}\left(\int \left[\Omega+k_\rho V\left(1+\frac{{\rm d}\rho}{{\rm d}\tau}\right)\right]{\rm d}\tau\right)=0.
\end{equation}

A zero imaginary part implies the absence of poles on the soft mode's trajectory. In fact, all potential poles proportional to $\Omega/N$ cancel exactly in the soft regime. This result is not surprising, since soft modes cross the horizon smoothly and do not show any peeling behavior that is needed for the Hawking effect. In previous articles \cite{ber13,fdp}, calculations have been performed within the EFB frame using the radial vector $\zeta$ and momentum $k_r=k\cdot\zeta$ which lead to the same result. However, as we will discover while studying the hard regime, the system bears a pathology which is cured only in the preferred frame by being seemingly absent in the soft regime.

\subsection{Hard Regime}
In contrast to the soft regime, the hard regime is characterized by modes that peel off the universal horizon instead of crossing it. As such, hard modes develop very high frequencies at the horizon, similar to the modes that constitute the relativistic Hawking effect. In other words, towards the horizon they experience a blue-shift beyond the Lorentz-breaking scale such that $\Lambda$ becomes extremely small when compared to the momentum. Mathematically, one might be tempted to take the limit $\Lambda\to0$, however the physically meaningful limit requires to keep track of the divergent dependence of $\omega$ as well. In this limit, \eqref{hjeq} becomes \eqref{hard} which is solved perturbatively to order $\mathcal{O}(N)$ through \cite{ding16}
\begin{equation}\label{pol}
k_\rho\simeq\frac{\Lambda}{V^{\nu}N^{\nu}}+\frac{V}{n-V^2}\Omega+\mathcal{O}(N),\quad\mbox{with}\quad\nu=\frac{1}{n-1}.
\end{equation}
The highest pole is proportional to $\Lambda$ and the next-to-leading order term yields a constant close to the universal horizon. We can insert the above form of $k_\rho$ into \eqref{ee} to obtain a formula for the preferred energy $\omega$. However, the second contribution, although seemingly subdominant, must not be ignored because it introduces a pole of order $1/N$ in $\omega$ and as such contributes to the prefactor in front of $\Omega$, indeed
\begin{equation}\label{kleinomega}
\omega\simeq\frac{nV^2}{n-V^2}\frac{\Omega}{N}+\frac{\Lambda V^{n\nu}}{N^{n\nu}}+\mathcal{O}(N^0),
\end{equation}
Note that the above prefactor, that we obtained naturally by working in the preferred frame, was instead introduced as a synchronization factor in the previous literature~\cite{fdp}. 

Obviously, the limit $\Lambda\to0$ yields, considering only the highest order of divergence, $k_\rho=0$. However this creates a tension with \eqref{ee} when evaluated at the universal horizon - unless we consider a mode with vanishing Killing energy. In general, the pole structure becomes complicated because of the fractional power $N^\nu$. Only for the case $\nu=1$ which is $n=2$, this pole is integrable because the branch-cut is absent\footnote{This can be derived from the fact that $1/N^2$ is a meromorphic function with double pole at $N=0$ with a residue of zero. From the residue theorem follows then that any circle around $N=0$ will not contribute to the contour integral. This is in agreement with \cite{ding16}.}. However, in general this pole leads to an infinite, irremovable contribution. If one wants to view this term as a chemical potential like in \cite{ding16}, its interpretation would be that at each instant of time infinitely many particles are emitted at the universal horizon. We think that this result closes the debate about which vacuum has to be used in this system: whenever Lorentz-breaking features dominates, the vacuum is fixed by the consistency of the theory itself, i.e. the only sensible vacuum is given by the preferred frame's observer. 
Hence, for this frame, we find for the principal function

\begin{equation}\label{pfpref}
\mathcal{S}_0=\int \left(\frac{nV^2}{n-V^2}\frac{\Omega}{N}+\frac{\Lambda V^{n\nu}}{N^{n\nu}}\right)N\mbox{d}\tau-\int\left(\frac{\Lambda}{N^{n\nu}V^{\nu}}+\frac{V}{(n-V^2)}\frac{\Omega}{N}\right)V\mbox{d}\rho
\end{equation}

The above expression shows that $\mathcal{S}_0$ is characterized by poles of different orders. For all cases $n\in\mathbb{N}\backslash\{2\}$ the pole $N^{n\nu}$ will be fractional. However, this pole cancels exactly between both integrals when the principal function is correctly evaluated along the tunneling path $\rho(\tau)$ using \eqref{traj}. Doing so we find
\begin{equation}\label{pftrajpref}
\mathcal{S}_0=\int\left(\frac{\Lambda V^{n\nu}}{N^{\nu}}\left(1-\frac{{\rm d}\rho}{{\rm d}\tau}\right)+\frac{V^2}{n-V^2}\Omega\left(n-\frac{{\rm d}\rho}{{\rm d}\tau}\right)\right)\mbox{d}\tau
\end{equation}
Our result depends explicitly on the signal speed on our trajectory d$\rho/$d$\tau$. For the hard modes, we use \eqref{pol} and \eqref{kleinomega} as well as $V\to1$ at $r=r_{\rm UH}$ to determine the signal speed at the universal horizon to be  
\begin{equation}
\frac{\mbox{d}\rho}{\mbox{d}\tau}=1+\mathcal{O}(N^\nu).\label{traj}
\end{equation}
Note, that this condition removes the intractable pole in the first term of expression \eqref{pftrajpref}.

As a next step, we recall the relation between the preferred time and the lapse function in proximity to the universal horizon \eqref{taulapse} and evaluate the principal function on the tunneling path.
By plugging \eqref{traj} into \eqref{pftrajpref}, not only the first pole cancels exactly in the considered limit, but also the second term simplifies (remember $V\to 1$ at the universal horizon) and we are left solely with a term being proportional to the Killing frequency so that
\begin{equation}\label{imteil}
\mbox{Im}(\mathcal{S}_0)=\mbox{Im}\left(\oint\Omega\mbox{d}\tau\right)=\mbox{Im}\left(\oint\frac{\Omega}{2\kappa_{\rm UH}}\frac{\mbox{d}N}{N}\right)=\frac{\pi\Omega}{2\kappa_{\rm UH}}
\end{equation}
Similarly to the relativistic black hole, we find a simple pole in the integration. Such poles lead to imaginary parts, as can be deduced by the Sokhotski-Plemelj theorem \cite{hormander,gelfand} for a function of form $f(x)/(x-x_o)$ after a complexification around the pole $x_o$
\begin{equation}\label{spt}
\lim_{\varepsilon\to0}\int_{a}^{b}\frac{f(x)}{x-x_o\pm i\varepsilon}\mbox{d}x=\mp i\pi f(x_o)+\mathcal{P}\left(\int_a^b\frac{f(x)}{x-x_o}\mbox{d}x\right)
\end{equation}
with $\mathcal{P}$ denoting the Cauchy principal value; the sign of the complexification depends on the specifics of the pole. Since the imaginary part in \eqref{imteil} is directly proportional to $\Omega$, our observer can compare it with the Boltzmann factor \eqref{boltz} and read off the universal temperature of the universal horizon to be
\begin{equation}\label{temuh}
T_{\rm UH}=\frac{\kappa_{\rm UH}}{\pi}.
\end{equation}

Let us take a moment to interpret our result: the temperature is universal, i.e. similar to the relativistic Hawking effect, just with a different surface gravity that is now governed by the universal horizon. This is insofar logical as within the ultralocal approximation around the horizon, the field behaves as a wave, thus, its speed is given by the speed of the wave front, which is arbitrarily large independently of the dispersion relations as long as this is Lorentz violating. This can be understood by taking into account the immense blue-shift that a wave package experiences when traced back to the horizon. Note that albeit this is a good description of the particle across the universal horizon, it is unsuitable to describe particles at the Killing horizon, given that Lorentz-breaking modes at most linger for a while there before crossing it. In that case, the trajectory is determined by the group velocity 
\begin{equation}
\frac{\mbox{d}\rho}{\mbox{d}\tau}=\left(\frac{{\rm d}k_\rho}{{\rm d}\omega}\right)^{-1}
\end{equation}
rather than the velocity of the wave front \eqref{traj}. Such group velocity depends on the highest order of Lorentz violation (which is \emph{a priori} different for different dispersion relations) while a purely monochromatic s-wave is ignorant to this. 

In the previous literature \cite{mi15}, some results involving the bounded mode $\phi_{\rm b}$ as the Hawking partner of $\phi_{\rm out}$ found a temperature governed by the Killing horizon. In the end, there are two separate processes that may contribute to the Hawking effect here. One is clearly the particle creation at the universal horizon involving $\phi_{\rm out}$ and $\phi_{\rm p}$, and another effective one between $\phi_{\rm out}$ and $\phi_{\rm b}$ within the ergoregion between the universal and Killing horizons. As the first one is reminiscent to the relativistic Hawking process, the latter resembles particle production in analog gravity \cite{fin12}. This mode might also feature a Penrose process alike phenomenon when scattering with $\phi_{\rm in}$. However, a deeper investigation of the phenomenology involving $\phi_{\rm b}$ is \emph{de rigeur}.  
\section{Discussion}

In this article we reviewed and revised the tunneling formalism for universal horizons within the framework of Lorentz-breaking theories. We showed that akin to the relativistic process, the temperature is governed by the horizon that introduces a causal separation, i.e. here, the universal rather than the Killing horizon. This substantiates the universality of black hole - or horizon - thermodynamics as a fundamental feature of quantum field theory even in the shadow of Lorentz breaking. We find explicitly that the definition of the Hawking effect remains intact and matches the usual result holding in horizon thermodynamics and consistency of quantum theory as shown in \cite{gia20}. Moreover, from the intricate mode structure originating from the abundance of higher derivatives, we concluded that only those modes, namely the hard modes, which peel off the horizon contribute to the Hawking effect. In this sense, the Hawking effect in Lorentz breaking gravity operates analogously to the relativistic case. However, due to the non-linear dispersion relation, we expect a later modification because the group velocity depends on the highest power in the Lorentz violation, thus affecting the signal speed.

Throughout our investigation we made use of two concepts that deserve particular attention: first the question of the tunneling path, and second the definition of the observer. Both concepts are \emph{a priori} not intertwined. However, the absence of Lorentz symmetry introduces a preferred physical frame which logically connects path and observer. For the theory to be well-defined, we must remain in the preferred frame, which automatically fixes our vacuum up to foliation preserving diffeomorphisms and the isometries of space-time itself. In this sense, the only family of vacua not plagued by Ostrogradskiy ghosts is the one where time aligns with the aether time. This reflects immediately back on the observer, who also needs to be located within a stable vacuum, i.e. one within the family of aether vacua. Whenever we are working in a different vacuum state, a non-removable pole occurs for the high-energy modes leading to an instantaneous creation of infinitely many particles. Hence, the quantum fields always sense the Lifshitz term in the action and must be evaluated within the preferred frame in order to yield a consistent quantum field theory.

We want to make here a remark on the consistency of QFT in this framework. As we saw for the relativistic case, the definition of the Hawking effect \eqref{ims0} is interwoven with a consistent probabilistic interpretation. Even under the absence of Lorentz symmetry, condition \eqref{ims0} holds. Another way of seeing this connection has been found by using analytic continuation through the universal horizon \cite{del22}. It can be shown that unless the lapse function admits certain properties, e.g. being $\mathcal{C}^2$ at the horizon, this will lead to a contradiction with the probabilistic interpretation of quantum theory. For those solutions (also in the relativistic case), the horizon temperature vanishes, which again fortifies the connection between thermodynamics and consistent quantum theories.

Although the system under consideration is additionally endowed with a Killing horizon, thermodynamics are ruled by the universal horizon. At least for the high energy limit, the Killing horizon does not play an essential role as argued in \cite{fdp}, although its low energy fate is under current investigation. For modes with momenta smaller than the Lorentz-breaking scale, the relativistic Hawking effect should be recovered. In these cases, we would expect a partnering between the bounded $\phi_{\rm b}$ and the outgoing mode $\phi_{\rm out}$. Note that due to the aforementioned negative Killing energy of the soft bounded mode seen  as the continuation of its hard counterpart, energy conservation for this partial relativistic Hawking effect is possible as in the standard case. 

The combination between $\phi_{\rm b}$ and $\phi_{\rm in}$ instead might feature the equivalent of a Penrose process, which enables the possibility of superradiance even in static black holes when involving an ingoing mode, as observed numerically in \cite{Oshita:2021onq}. Owing to the peculiar mode structure, negative energy modes living beyond the Killing horizon can transfer energy to ingoing modes, turning them into outgoing modes while itself falling into the singularity. Through these modes, a mining of energy from the region between universal and Killing horizon becomes possible.

Our results present a harbinger for quantum gravity in the following sense: whatever process forms the horizon has to support thermal quantum effects provided that the limit where QFT in curved space-times applies can be dynamically achieved. Consequently, black hole thermodynamics is a resilient process perpetuated at causal barriers. Even in the absence of Lorentz symmetry, the mechanism behind the Hawking effect works universally, i.e. exactly identical to the relativistic case, deeming the process to be fundamental.

\ack

Thank you to the organizers of the conference ``Avenues in Quantum field Theory in Curved Spacetimes 2022" held in Genoa especially Vincenzo Vitagliano for his devotion during the conference. MS wants to thank the organizers for getting the opportunity to present this work within the magnificent and breathtaking ambience of the lecture hall from the XVIIth century located within the Unesco world heritage of Strada Balbi. The work of F. D. P., S. L., and M. S. has been supported by the Italian Ministry of Education and Scientific Research (MIUR) under the Grant PRIN MIUR 2017-MB8AEZ. The work of M.H.-V. has been supported by the Spanish State Research Agency MCIN/AEI/10.13039/501100011033 and by the EU NextGenerationEU/PRTR funds, under Grant No. IJC2020-045126-I. IFAE is partially funded by the CERCA program of the Generalitat de Catalunya
\section*{References}
\bibliographystyle{iopart-num.bst}
\bibliography{cptunnel.bib}

\providecommand{\newblock}{}
\begin{thebibliography}{10}
\expandafter\ifx\csname url\endcsname\relax
  \def\url#1{{\tt #1}}\fi
\expandafter\ifx\csname urlprefix\endcsname\relax\def\urlprefix{URL }\fi
\providecommand{\eprint}[2][]{\url{#2}}

\bibitem{haw75}
Hawking S~W 1975 Particle creation by black holes {\em Euclidean quantum
  gravity\/} (World Scientific) pp 167--188

\bibitem{har83}
Hartle J~B and Hawking S~W 1983 Wave function of the universe {\em Euclidean
  quantum gravity\/} (World Scientific) pp 310--325

\bibitem{gib93}
Gibbons G~W and Hawking S~W 1993 Cosmological event horizons, thermodynamics,
  and particle creation {\em Euclidean quantum gravity\/} (World Scientific) pp
  281--294

\bibitem{hay94}
Hayward S~A 1994 {\em Physical Review D\/} {\bf 49} 6467

\bibitem{ash99}
Ashtekar A, Beetle C and Fairhurst S 1999 {\em Classical and Quantum Gravity\/}
  {\bf 16} L1

\bibitem{ash00}
Ashtekar A, Beetle C, Dreyer O, Fairhurst S, Krishnan B, Lewandowski J and
  Wi{\'s}niewski J 2000 {\em Physical Review Letters\/} {\bf 85} 3564

\bibitem{ash03}
Ashtekar A and Krishnan B 2003 {\em Physical Review D\/} {\bf 68} 104030

\bibitem{Barcelo:2010xk}
Barcelo C, Liberati S, Sonego S and Visser M 2011 {\em JHEP\/} {\bf 02} 003
  (\textit{Preprint} \eprint{1011.5911})

\bibitem{Barcelo:2010pj}
Barcelo C, Liberati S, Sonego S and Visser M 2011 {\em Phys. Rev. D\/} {\bf 83}
  041501 (\textit{Preprint} \eprint{1011.5593})

\bibitem{par00}
Parikh M~K and Wilczek F 2000 {\em Phys. Rev. Lett.\/} {\bf 85} 5042--5045
  (\textit{Preprint} \eprint{hep-th/9907001})

\bibitem{apd00}
Srinivasan K and Padmanabhan T 1999 {\em Phys. Rev. D\/} {\bf 60} 024007
  (\textit{Preprint} \eprint{gr-qc/9812028})

\bibitem{mas00}
Massar S and Parentani R 2000 {\em Nuclear Physics B\/} {\bf 575} 333--356

\bibitem{pad02}
Sriramkumar L and Padmanabhan T 2002 {\em International Journal of Modern
  Physics D\/} {\bf 11} 1--34

\bibitem{van08}
Di~Criscienzo R and Vanzo L 2008 {\em EPL (Europhysics Letters)\/} {\bf 82}
  60001

\bibitem{di09}
Di~Criscienzo R, Hayward S~A, Nadalini M, Vanzo L and Zerbini S 2009 {\em
  Classical and Quantum Gravity\/} {\bf 27} 015006

\bibitem{van11}
Vanzo L, Acquaviva G and Di~Criscienzo R 2011 {\em Classical and Quantum
  Gravity\/} {\bf 28} 183001

\bibitem{far14}
Faraoni V and Vitagliano V 2014 {\em Physical Review D\/} {\bf 89} 064015

\bibitem{gia20}
Giavoni C and Schneider M 2020 {\em Class. Quant. Grav.\/} {\bf 37} 215020
  (\textit{Preprint} \eprint{2003.11095})

\bibitem{par04}
Parikh M 2004 {\em International Journal of Modern Physics D\/} {\bf 13}
  2351--2354

\bibitem{sen07}
Senovilla J~M 2007 {\em Classical and Quantum Gravity\/} {\bf 24} 3091

\bibitem{Horava:2009uw}
Ho\v{r}ava P 2009 {\em Phys. Rev. D\/} {\bf 79} 084008 (\textit{Preprint}
  \eprint{0901.3775})

\bibitem{berg12}
Berglund P, Bhattacharyya J and Mattingly D 2012 {\em Physical Review D\/} {\bf
  85} 124019

\bibitem{Liberati:2013xla}
Liberati S 2013 {\em Class. Quant. Grav.\/} {\bf 30} 133001 (\textit{Preprint}
  \eprint{1304.5795})

\bibitem{Addazi:2021xuf}
Addazi A {\em et~al.\/} 2022 {\em Prog. Part. Nucl. Phys.\/} {\bf 125} 103948
  (\textit{Preprint} \eprint{2111.05659})

\bibitem{ber13}
Berglund P, Bhattacharyya J and Mattingly D 2013 {\em Physical review
  letters\/} {\bf 110} 071301

\bibitem{mi15}
Michel F and Parentani R 2015 {\em Physical Review D\/} {\bf 91} 124049

\bibitem{he21}
Herrero-Valea M, Liberati S and Santos-Garcia R 2021 {\em Journal of High
  Energy Physics\/} {\bf 2021} 1--32

\bibitem{fdp}
Del~Porro F, Herrero-Valea M, Liberati S and Schneider M 2022 {\em Phys. Rev.
  D\/} {\bf 106} 064055 (\textit{Preprint} \eprint{2207.08848})

\bibitem{hay09}
Hayward S~A, Di~Criscienzo R, Nadalini M, Vanzo L and Zerbini S 2009 {\em
  Classical and Quantum Gravity\/} {\bf 26} 062001

\bibitem{sen15}
Senovilla J~M and Torres R 2015 {\em Classical and quantum gravity\/} {\bf 32}
  085004

\bibitem{Jacobson:2003vx}
Jacobson T 2003 {Introduction to quantum fields in curved space-time and the
  Hawking effect} {\em {School on Quantum Gravity}\/} pp 39--89
  (\textit{Preprint} \eprint{gr-qc/0308048})

\bibitem{mor12}
Moretti V and Pinamonti N 2012 {\em Communications in Mathematical Physics\/}
  {\bf 309} 295--311

\bibitem{kur21}
Kurpicz F, Pinamonti N and Verch R 2021 {\em Letters in Mathematical Physics\/}
  {\bf 111} 1--44

\bibitem{Anselmi:2007ri}
Anselmi D and Halat M 2007 {\em Phys. Rev. D\/} {\bf 76} 125011
  (\textit{Preprint} \eprint{0707.2480})

\bibitem{Barvinsky:2015kil}
Barvinsky A~O, Blas D, Herrero-Valea M, Sibiryakov S~M and Steinwachs C~F 2016
  {\em Phys. Rev. D\/} {\bf 93} 064022 (\textit{Preprint} \eprint{1512.02250})

\bibitem{Barvinsky:2017kob}
Barvinsky A~O, Blas D, Herrero-Valea M, Sibiryakov S~M and Steinwachs C~F 2017
  {\em Phys. Rev. Lett.\/} {\bf 119} 211301 (\textit{Preprint}
  \eprint{1706.06809})

\bibitem{Barvinsky:2019rwn}
Barvinsky A~O, Herrero-Valea M and Sibiryakov S~M 2019 {\em Phys. Rev. D\/}
  {\bf 100} 026012 (\textit{Preprint} \eprint{1905.03798})

\bibitem{Bellorin:2022qeu}
Bellorin J, Borquez C and Droguett B 2022 {\em Phys. Rev. D\/} {\bf 106} 044055
  (\textit{Preprint} \eprint{2207.08938})

\bibitem{Barvinsky:2023mrv}
Barvinsky A~O 2023  (\textit{Preprint} \eprint{2301.13580})

\bibitem{Jacobson:2000xp}
Jacobson T and Mattingly D 2001 {\em Phys. Rev. D\/} {\bf 64} 024028
  (\textit{Preprint} \eprint{gr-qc/0007031})

\bibitem{Mattingly:2005re}
Mattingly D 2005 {\em Living Rev. Rel.\/} {\bf 8} 5 (\textit{Preprint}
  \eprint{gr-qc/0502097})

\bibitem{yagi14}
Yagi K, Blas D, Barausse E and Yunes N 2014 {\em Physical Review D\/} {\bf 89}
  084067

\bibitem{Gupta:2021vdj}
Gupta T, Herrero-Valea M, Blas D, Barausse E, Cornish N, Yagi K and Yunes N
  2021 {\em Class. Quant. Grav.\/} {\bf 38} 195003 (\textit{Preprint}
  \eprint{2104.04596})

\bibitem{eli06}
Eling C and Jacobson T 2006 {\em Classical and Quantum Gravity\/} {\bf 23}
  5643--5660 \urlprefix\url{https://doi.org/10.1088/0264-9381/23/18/009}

\bibitem{bar11}
Barausse E, Jacobson T and Sotiriou T~P 2011 {\em Phys. Rev. D\/} {\bf 83}
  124043 (\textit{Preprint} \eprint{1104.2889})

\bibitem{Pospelov:2010mp}
Pospelov M and Shang Y 2012 {\em Phys. Rev. D\/} {\bf 85} 105001
  (\textit{Preprint} \eprint{1010.5249})

\bibitem{Blas:2010hb}
Blas D, Pujolas O and Sibiryakov S 2011 {\em JHEP\/} {\bf 04} 018
  (\textit{Preprint} \eprint{1007.3503})

\bibitem{Bhattacharyya:2015gwa}
Bhattacharyya J, Colombo M and Sotiriou T~P 2016 {\em Class. Quant. Grav.\/}
  {\bf 33} 235003 (\textit{Preprint} \eprint{1509.01558})

\bibitem{Carballo-Rubio:2020ttr}
Carballo-Rubio R, Di~Filippo F, Liberati S and Visser M 2020 {\em JHEP\/} {\bf
  12} 055 (\textit{Preprint} \eprint{2005.08533})

\bibitem{Carballo-Rubio:2021wjq}
Carballo-Rubio R, Di~Filippo F, Liberati S and Visser M 2022 {\em JHEP\/} {\bf
  02} 122 (\textit{Preprint} \eprint{2111.03113})

\bibitem{bhatt}
Bhattacharyya J 2013 {\em Aspects of holography in Lorentz-violating gravity\/}
  Ph.D. thesis

\bibitem{cro14}
Cropp B, Liberati S, Mohd A and Visser M 2014 {\em Physical Review D\/} {\bf
  89} 064061

\bibitem{del22}
Del~Porro F, Herrero-Valea M, Liberati S and Schneider M 2022 {\em Phys. Rev.
  D\/} {\bf 105} 104009 (\textit{Preprint} \eprint{2201.03584})

\bibitem{cro13}
Cropp B, Liberati S and Visser M 2013 {\em Classical and Quantum Gravity\/}
  {\bf 30} 125001

\bibitem{ding16}
Ding C, Wang A, Wang X and Zhu T 2016 {\em Nuclear Physics B\/} {\bf 913}
  694--715

\bibitem{hormander}
H{\"o}rmander L 2015 {\em The analysis of linear partial differential operators
  I: Distribution theory and Fourier analysis\/} (Springer)

\bibitem{gelfand}
Gelfand I and Shilov G 1964 {\em Generalized Functions\/} vol~1 (Academic
  Press)

\bibitem{fin12}
Finazzi S and Parentani R 2012 {\em Physical Review D\/} {\bf 85} 124027

\bibitem{Oshita:2021onq}
Oshita N, Afshordi N and Mukohyama S 2021 {\em JCAP\/} {\bf 05} 005
  (\textit{Preprint} \eprint{2102.01741})

\end{thebibliography}

\end{document}